\documentclass[conference]{IEEEtran}
\IEEEoverridecommandlockouts
\usepackage{cite}
\usepackage{amsmath,amssymb,amsfonts}
\usepackage{algorithmic}
\usepackage{graphicx}
\usepackage{textcomp}
\usepackage{xcolor}
\usepackage{booktabs}
\usepackage{hyperref}
\usepackage{amsthm}
\usepackage{caption}
\usepackage{multirow}

\def\BibTeX{{\rm B\kern-.05em{\sc i\kern-.025em b}\kern-.08em
    T\kern-.1667em\lower.7ex\hbox{E}\kern-.125emX}}
\begin{document}

\title{Improving Out-of-Domain Audio Deepfake Detection via Layer Selection and Fusion of SSL-Based Countermeasures

%\thanks{anonymous funding agency}

}

\author{\IEEEauthorblockN{1\textsuperscript{st} Pierre Serrano}
 \IEEEauthorblockA{\textit{Inria Défense\&Sécurité, LR2}\\
 pierre.serrano@inria.fr}
 \and
 \IEEEauthorblockN{2\textsuperscript{nd} Raphaël Duroselle}
 \IEEEauthorblockA{\textit{Inria Défense\&Sécurité, LR2}}
 \and
 \IEEEauthorblockN{3\textsuperscript{rd} Florian Angulo}
 \IEEEauthorblockA{\textit{Inria Défense\&Sécurité, LR2}}
 \and
 \IEEEauthorblockN{4\textsuperscript{rd} Jean-François Bonastre}
 \IEEEauthorblockA{\textit{Inria Défense\&Sécurité, LR2}}
 \and
 \IEEEauthorblockN{5\textsuperscript{th} Olivier Boeffard}
 \IEEEauthorblockA{\textit{Inria Défense\&Sécurité, LR2}}
 \and
 }

\maketitle

\begin{abstract}

Audio deepfake detection systems based on frozen pre-trained self-supervised learning (SSL) encoders show a high level of performance when combined with layer-weighted pooling methods, such as multi-head factorized attentive pooling (MHFA). However, they still struggle to generalize to out-of-domain (OOD) conditions. We tackle this problem by studying the behavior of six different pre-trained SSLs, on four different test corpora. We perform a layer-by-layer analysis to determine which layers contribute most. Next, we study the pooling head, comparing a strategy based on a single layer with automatic selection via MHFA. We observed that selecting the best layer gave very good results, while reducing system parameters by up to 80\%. A wide variation in performance as a function of test corpus and SSL model is also observed, showing that the pre-training strategy of the encoder plays a role. Finally, score-level fusion of several encoders improved generalization to OOD attacks.

\end{abstract}


\section{Introduction}

Recent advances in audio deepfake detection have been driven by self-supervised learning (SSL) models, which combine pre-trained encoders with lightweight classification heads \cite{wang24_ASVspoof}.
These pipelines significantly outperform previous methods, with Equal Error Rates (EERs) on challenging datasets such as InTheWild (ITW) dropping from over 30\% \cite{muller_does_2024} to below 10\% in recent studies \cite{doan_trident_2024, zhang_audio_2024}.

The best-performing systems often rely on SSL encoders such as wav2vec2.0 or WavLM. Some have achieved strong results even with frozen backbones \cite{tak2022automatic, martin-donas_exploring_2024},
which considerably simplifies the whole development process. %, ASVspoof2025.
Despite these improvements, the detection of unseen attacks in out-of-domain (OOD) conditions remains a major challenge. For instance, EERs on ITW or multilingual datasets such as MLAAD can still exceed 20\% \cite{pascu2024towards,kulkarni2024exploring}.
This raises open questions about the behavior of these SSL models, including the influence of pre-training objectives and model architectures on cross-domain generalization. %yamaguchi2023

While several studies have compared the performance of individual SSL models \cite{pascu2024towards,kheir_comprehensive_2025}, few have investigated their fusion \cite{kulkarni2024exploring}.
We believe that combining models with diverse pre-training objectives could enhance OOD generalization by leveraging complementary strengths.
However, combining full SSL pipelines leads to substantial increases in inference cost.
A promising alternative lies in identifying and exploiting the most informative internal layers following early exit strategy \cite{pimentel2024efficient}.
Indeed, prior work has shown that intermediate layers can encode task-relevant features \cite{pasad_layer-wise_2022}, which opens the door to more efficient and scalable fusion strategies.

\begin{table*}[th]
\caption{Training, validation and evaluation datasets.}
\label{tab:datasets}
%\vspace{-2mm}
\centerline{\begin{tabular}{|c|c|c|c|c|c|c|}
\hline
Dataset & Usage & \# speakers & \# bona & \# spoof & \# attacks & \# languages \\
\hline
ASVspoof5-train-train & training & 327 & 15156 & 130440  & 8 & 1 \\
ASVspoof5-train-dev & validation  & 73 & 3613 & 32858  & 8 & 1  \\
ASVspoof5-dev & test and fusion & 785 & 31334 & 109616 &  8 & 1 \\
ASVspoof5-eval & test & 737 & 138688 & 542086  &  16 & 1 \\
InTheWild & test & 54 & 19963 & 11816 &  - & 1 \\
MLAAD v5 + M-AILABS & test & - & 41000  & 41000  & 7 & 7 \\
LlamaPartialSpoof & test & - &  10573 &  65655 & 6 & 1 \\
\hline
\end{tabular}
}
\end{table*}

%jf
In this work, we address the following research questions:
\begin{enumerate}
    \item Depending on the SSL model, which layers generalize best to out-of-domain conditions?
    \item Can state-of-the-art performance be achieved in OOD settings with frozen backbones and limited training data?
    \item Should different SSL encoders be preferred depending on the targetted test condition and could we gain in OOD generalization by combining different encoders?
\end{enumerate}

To answer these questions, we explore how to optimally exploit and combine frozen SSL systems for audio deepfake detection under OOD conditions. 
To this end, we propose a methodology that identifies the most informative layers for deepfake detection. 
Our experiments involve six SSL backbones, varying in size and pre-training tasks.
We freeze all encoders to isolate the effect of their representations, and train classification heads using only the ASVspoof5 train set,
which includes eight types of attacks. Evaluation is performed on unseen datasets with different attack generation methods.
We compare manual and automatic layer selection approaches. We also evaluate the potential of score-level fusion to improve generalization while maintaining low computational cost.

The main contribution of this paper is a systematic layer-wise analysis across SSL models, 
revealing that intermediate layers consistently provide the most relevant features for audio deepfake detection. 
Selecting a MHFA optimal layer allows us to approach the performance of more complex pooling strategies like LWA,
while significantly reducing the number of parameters. Moreover, we demonstrate that OOD generalization varies strongly across SSL models. 
Leveraging their complementarity through score-level fusion of simple classifiers yields strong results across several datasets.
Notably, we achieve state-of-the-art performance in OOD conditions despite using limited training data and no data augmentation.

This paper is organized as follows. Section~\ref{sec:section2} presents our methodology and classification heads. Section~\ref{sec:section3} details the training protocol and evaluation setup. Section~\ref{sec:section4} reports and discusses our results. Section~\ref{sec:section5} concludes and outlines future directions.

\section{Analysis of self-supervised audio encoders}
\label{sec:section2}

Our objective is to assess how effectively SSL representations, particularly from specific layers, can be leveraged for audio deepfake detection.
For this study, the SSL backbones are frozen to investigate how the pre-trained representations, without any additional fine-tuning, perform in the task of deepfake detection.
Building on recent work that investigates the in-domain performance of classifiers using different layers of SSL encoders \cite{kheir_comprehensive_2025}, we focus in this work on generalization to OOD conditions.

The proposed methodology consists in training several classification heads on top of the frozen encoders.  In the following sections, we outline the two classification heads used in our experiments.

The output of the $l$-th Transformer layer of the encoder is $Z_{l} \in \mathbb{R}^{T \times F}$ with $l \in \lbrace 1,...,L \rbrace$, where $L$ denotes the total number of Transformer layers in pre-trained model, $T$ the number of temporal frames, and $F$ the features dimension.
Both classification heads are constituted of a pooling layer that takes as input the activations $Z$ of the Transformer layers and produces a fixed-dimensional embedding $\mathbf{e} \in \mathbb{R}^{D}$, and of a final linear layer for the spoofing binary detection task.  The total number of trainable parameters associated with the classification heads is negligible compared to the number of SSL parameters, ranging from 99K to 330K.

\subsection{Mean pooling (MP)}

To evaluate the contribution of each layer's output to deepfake detection, we attach a basic MP head to the output of each hidden layer in the model. This step is essential to propose an optimal single-layer model for each SSL backbone, based on the layer's ability to capture crucial features.

The MP operation reduces the dimensionality of the feature vectors by averaging the values across the temporal frames. First, the frame-level feature vectors are projected to dimension $D$ with a linear layer. Then, the MP operation computes the average of the feature vectors across the temporal dimension to produce the final mean-pooled vector $\mathbf{e} \in \mathbb{R}^{D}$.
%

%
%
% \subsection{Layer-weighted average (LWA)}
%
% Next, we conduct experiments to evaluate the added value of LWA. This approach isolates the effect of the layer selection mechanism, allowing us to assess the model’s ability to weight and aggregate layer outputs effectively.
% In LWA, trainable weights $w_l$ are assigned to each layer $l$, enabling the model to learn a weighted average of the outputs from all layers to generate frame-level features $\mathbf{O} \in \mathbb{R}^{T \times F}$. Then a Mean pooling is applied.

\subsection{Multi-head factorized attentive pooling (MHFA)}

We then evaluate the added value of the combination of the activations of all layers of the SSL encoder. We use the MHFA layer \cite{peng_attention-based_2023} that combines LWA with an attention mechanism.
%We then attach a MHFA head to each SSL backbone. This mechanism combines LWA pooling with an attention mechanism \cite{peng_attention-based_2023}. The goal is to evaluate how MHFA, with its attention mechanism, enhances the model's ability to detect deepfakes under OOD conditions.

In MHFA, two sets of weights, $w^k$ and $w^v$, are used to aggregate layer-wise outputs and produce matrices of keys $\mathbf{K} \in \mathbb{R}^{T,D}$ and values $\mathbf{V} \in \mathbb{R}^{T,D}$ with $D$ the hidden size dimension. The matrices $S^k$ and $S^v$ are used to reduce dimension from $F$ to $D$.
\[
 K = (\sum_{l=1}^L w_l^k Z_l) S^k \quad  V = (\sum_{l=1}^L w_l^v Z_l) S^v
\]

The matrices $K$ and $V$ are then passed through an attention pooling mechanism detailed in \cite{peng_attention-based_2023}, that produces the embedding $\mathbf{e} \in \mathbb{R}^{D}$.

\section{Experimental protocols}
\label{sec:section3}

\subsection{Self-supervised pretrained backbones}

The models selected for this study cover a range of sizes and pre-training tasks, as commonly seen in the literature. A detailed list of these models and their characteristics is provided in Table \ref{tab:tab_model}.

%Wav2vec 2.0 Base and WavLM Base were among the models permitted in the ASVspoof5 challenge. WavLM Large and MMS represent evolutions of these models.

Wav2vec 2.0 Base \cite{baevski_wav2vec_2020} and WavLM Base \cite{chen_wavlm_2022} were widely used by participants to the ASVspoof5 challenge \cite{wang24_ASVspoof}. WavLM Large \cite{chen_wavlm_2022} and MMS \cite{pratap_scaling_2023} represent larger versions of these models, trained on larger datasets with more parameters. However, they may introduce bias due to an overlap between their training data and the challenge dataset. Wav2vec2-XLS-R (0.3B) \cite{babu2022xls} is a multilingual version of intermediate size of the Wav2vec 2.0 family of models, which have been demonstated very effective for audio deepfake detection \cite{tak2022automatic}.
All of these speech models are built on similar Transformer-based architectures, though they differ in their pre-training tasks and training datasets. Further details can be found in \cite{zaiem_speech_2023}.

Originally designed for tasks like audio captioning and sound event detection \cite{chen_beats_2022}, BEATs has recently shown impressive results in audio deepfake detection in a speaker aware setup~\cite{pianese_training-free_2024}. Based on these findings, we decided to evaluate BEATs.

\begin{table}[th]
 %\captionsetup{justification=centering}
  \caption{SSL audio encoders used as backbone}
  %\vspace{-2mm}
  \label{tab:tab_model}
  \centering
  %\footnotesize % Change la taille de la police à \scriptsize
  \begin{tabular}{c|c|c}
    \toprule
    \textbf{Model}   & \textbf{\# parameters} & \textbf{\# hidden layers} \\
    \midrule
     Wav2vec 2.0 Base \cite{baevski_wav2vec_2020}     & 94M  & 12  \\
     WavLM Base   \cite{chen_wavlm_2022}        & 94M  & 12  \\
     BEATs        \cite{chen_beats_2022}        & 94M  & 12  \\
     Wav2vec 2.0 XLS-R      \cite{babu2022xls}     & 315M     & 24  \\
     WavLM Large  \cite{chen_wavlm_2022}          & 317M & 24  \\
     MMS         \cite{pratap_scaling_2023}        & 1B  & 48  \\
    \bottomrule
  \end{tabular}
\end{table}

\begin{figure*}[!th]
  \centering
 \captionsetup{justification=centering}
  \includegraphics[width=\linewidth]{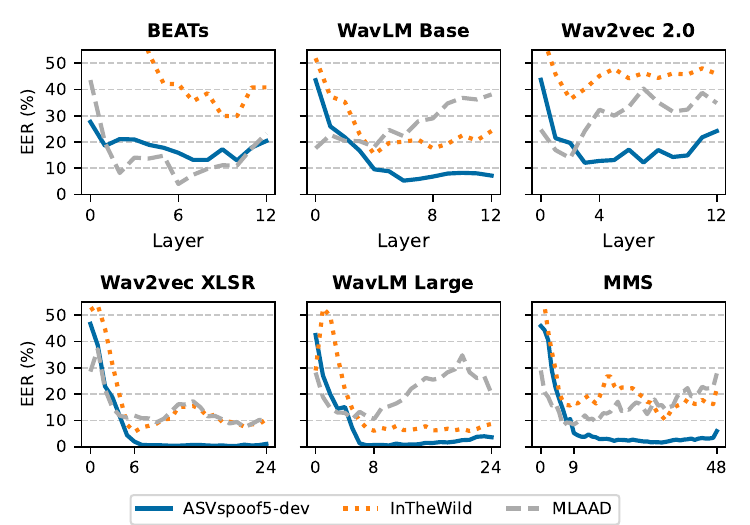}%layer_wise_analyse_INK2.pdf}
  %\vspace{-0.5cm}
	\caption{Layer-wise Equal Error Rate (EER) analysis for all SSL models. Each subplot corresponds to a different SSL model, and each line within a subplot represents the EER obtained on a specific evaluation dataset. A mean pooling classifier is trained independently on each layer of each model to assess its effectiveness in detecting deepfakes. The best single-layer models, discussed in Table \ref{tab:MHFA_results_2}, are indicated on the x-axis. Note that Layer 0 corresponds to the output of the convolutional layers of the models.  } %Dataset abbreviations: ASV5 = ASVspoof5-dev, ITW = InTheWild, MLAAD = MLAAD v5 + M-AILABS.
  \label{fig:layer_wise_analysis}
\end{figure*}

\begin{figure*}[!th]
  \centering
 \captionsetup{justification=centering}
  \includegraphics[width=\linewidth]{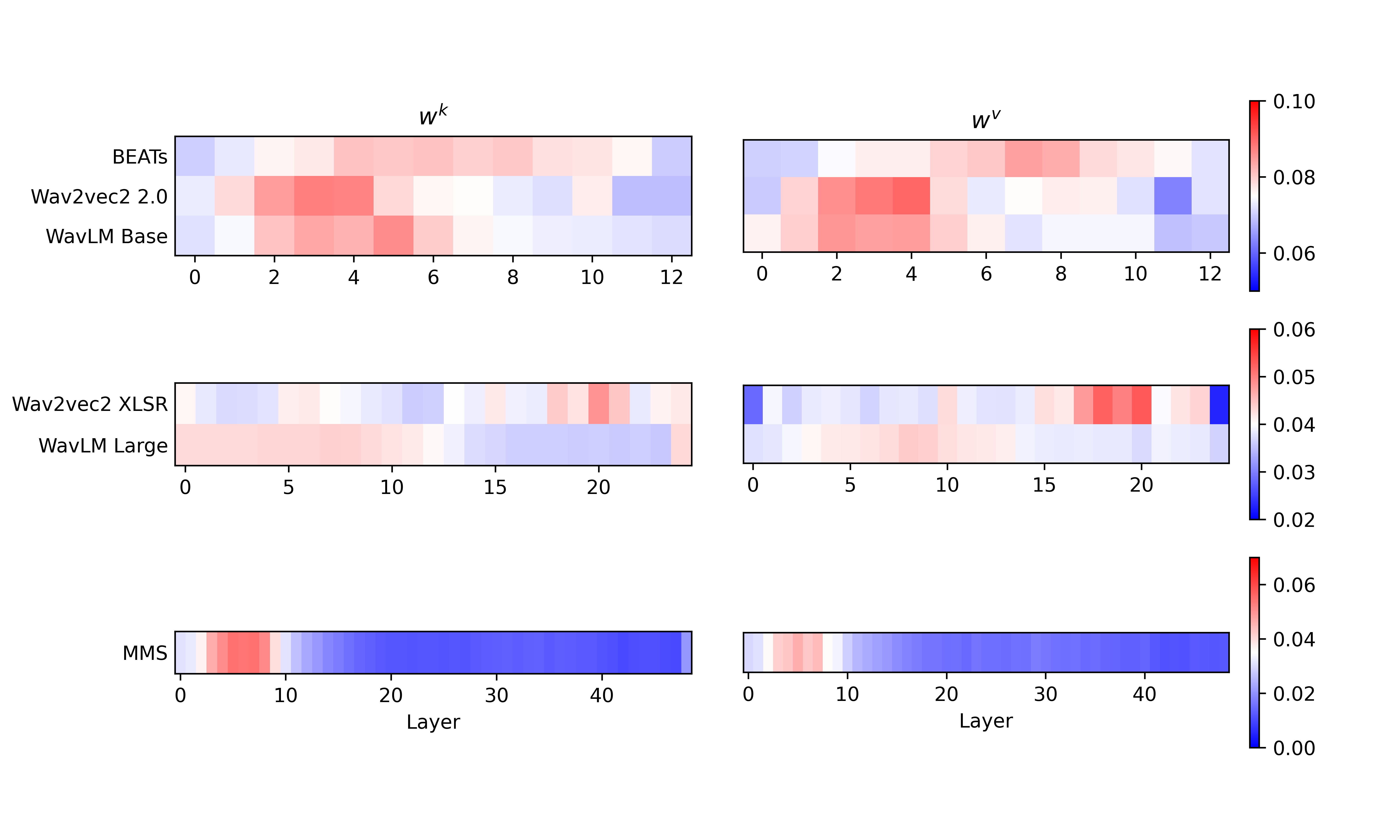}%layer_wise_analyse_INK2.pdf}
  \vspace{-0.5cm}
	\caption{MHFA weights for each layer and backbone. A different color scale is chosen for each size of model since the sum of the weights of all layers of a model is one.}
  \label{fig:MHFA_weights}
\end{figure*}

\subsection{Datasets}

The details of the datasets used in our study are provided in \autoref{tab:datasets}.
All classifiers are trained on the ASVspoof5 training corpus, which contains eight attacks generated using only four different text-to-speech systems.
We selected this reduced training corpus because it offers a challenging scenario for OOD generalization and it enables comparison with recent works in the ASVspoof5 challenge.
%We use the ASVspoof5 training corpus to train all models.
The ASVspoof5 training corpus is split into two subsets: ASVspoof5-train-train which contains 80\% of the speakers and is used for training, and ASVspoof5-train-dev, which contains the remaining ones for validation.

For evaluation, we consider multiple evaluation datasets to measure performance on OOD conditions:
\begin{itemize}
\item ASVspoof5-dev \cite{wang24_ASVspoof}: This dataset includes 8 unseen attacks, specifically generated according to the ASVspoof5 challenge's methodology.
\item ASVspoof5-eval \cite{wang24_ASVspoof}: the evaluation set of the ASVspoof5 challenge, that includes 16 unseen attacks, with adversarial attacks and codec augmentations.
\item InTheWild \cite{muller_does_2024}:  This dataset includes fake samples collected from social networks. The exact types of attacks are unknown, but it serves to evaluate the model's ability to generalize to various unseen attack types in an application context.
\item MLAAD v5 + M-AILABS \cite{muller_mlaad_2024}: MLAAD consists of TTS attacks and is multilingual. For the bonafide class, we include bonafide samples from the multilingual M-AILABS dataset as suggested by the MLAAD authors. The identity of the target speakers is not defined in MLAAD.
The test set is balanced by considering spoof utterances and bona fide samples in seven languages from the M-AILABS dataset, excluding English. This ensures an equal representation of both classes in these unseen languages, allowing us to evaluate the model's performance on languages that were not present during training.
Additionally, we remove the Griffin-Lim attack from MLAAD, as we do not consider it a deepfake in this dataset. A deepfake should involve a form of speech manipulation that enables semantic modification (i.e. making someone say something they didn’t say). Instead, this attack reconstructs the signal from bona fide magnitude spectrograms of the MAILABS dataset without altering the speaker identity or content.
\item LlamaPartialSpoof \cite{luong2025llamapartialspoof}: This dataset simulates a more challenging disinformation generation process designed from attackers’ perspective. We limit our evaluation to the full utterances subset (excluding partial fakes).
\end{itemize}

All datasets are comparable to the litterature, with the exception of MLAAD+M-AILABS where we define a custom subset to evaluate performance on unseen languages.

\subsection{Training setup \& evaluation metrics}

All training experiments are conducted using the same hyperparameters: hidden dimension $D$ of 128, 8 attention heads, batch size 128 and audio split into 3-second segments. The classification heads are trained on 2 NVIDIA A100 80GB GPUs. We use the Adam optimizer with betas respectively set to $0.90$ and $0.98$, $\varepsilon = 1^{-8}$, weight decay = $2e^{-6}$, and a constant learning rate of $1e^{-4}$.  Training continues for a minimum of 6 epochs, with additional epochs added if convergence has not been achieved. Checkpoints are selected based on the minimum validation loss. The evaluation metric used is the Equal Error Rate (EER).
No data augmentation is applied.

Even though all systems are trained with 3-second segments, the pooling mechanism allows the model to handle arbitrary duration at inference. We do not use any technique to adapt the models to variable durations. For inference, the maximum audio duration is limited to the first 30 seconds of each utterance.

The training converged extremely fast, reaching near-zero validation loss and EER, suggesting that the task is relatively simple. Despite the simplicity of the classifier attached to each layer, increasing its complexity when using MHFA does not alter the observed global behavior.

\subsection{Fusion of detectors}

With the goal of evaluating the complementarity of SSL encoders for OOD deepfake detection, we evaluate simple score-level combinations of systems.  Fusion is performed with a simple logistic regression \cite{kulkarni2024exploring} with the implementation of \cite{alumae2019taltech}.% with the implementation of \textit{https://github.com/alumae/sv\_score\_calibration}.

To train the logistic regression, we neeed a fusion corpus where systems reach non zero EERs. In practice in our experiments, that corresponds to a fusion corpus with unseen attacks. Nonetheless the choice of a specific set of attacks to learn the fusion model is critical, especially for generalization to unseen conditions.

Abiding by the ASVspoof5 challenge protocol, we train the fusion model on the corpus ASVspoof5-dev. This choice is not perfect but corresponds to unseen attacks, assumed closed to the training conditions of the systems. To  reduce the impact of this choice, we compare the fusion of several systems with logistic regression to the simple sum of calibrated log-likelihood ratio (calibration is done on the corpus ASVspoof5-dev).

\section{Experimental results}
\label{sec:section4}

%
% \begin{comment}
%
%
% \begin{table*}[htbp]
% \centering
% \caption{Equal Error Rate (EER) obtained for each backbone using three approaches: the best single-layer with mean pooling (BSL), layer-weighted average (LWA), and multi-head factorized attentive pooling (MHFA). The best layer number for the BSL approach is indicated in the column "BSL \#", selected manually as an example to demonstrate the potential of identifying a relevant layer for each model.}
% \label{tab:MHFA_results}
% %\vspace{-2mm}
% \centering
% \begin{tabular}{ll|rrrrrrrrrrrrrrr}
% \toprule
% & & \multicolumn{3}{c}{ASVspoof5-dev} & \multicolumn{3}{c}{InTheWild} & \multicolumn{3}{c}{MLAAD v5 + M-AILABS} & \multicolumn{3}{c}{ASVspoof5-eval} & \multicolumn{3}{c}{LlamaPartialSpoof} \\
% \cmidrule(lr){3-5} \cmidrule(lr){6-8} \cmidrule(lr){9-11}\cmidrule(lr){12-14}\cmidrule(lr){15-17}
% Backbone & BSL \# & BSL & LWA & MHFA & BSL & LWA & MHFA & BSL & LWA & MHFA& BSL & LWA & MHFA& BSL & LWA & MHFA \\
% \midrule
% BEATs       & 6 & 15.9 & 15.1 & 16.7 & 42.0 & 37.2 & 40.2 & 4.1  & 15.7 & 7.7 &&&&&& \\
% Wav2vec 2.0 & 8 & 19.6 & 13.3 & 10.8 & 36.3 & 40.5 & 32.3 & 13.9 & 22.3 & 20.2 &&&&&& \\
% WavLM Base  & 4 & 9.6  & 2.1  & 2.5  & 15.3 & 19.5 & 16.7 & 18.1 & 22.3 & 16.1 &&&&&& \\
% WavLM Large & 8 & 0.6  & 1.5  & 1.5  & 6.0  & 5.7  & 5.4  & 10.5 & 24.1 & 15.8&&&&&&  \\
% MMS         & 9 & 5.2  & 2.8  & 6.2  & 17.0 & 20.3 & 18.1 & 8.4 & 8.8  & 11.4 &&&&&& \\
% Wav2vec2-XLS-R         & 6 &  & &  & & &  & &  &  &&&&&& \\
% \bottomrule
% \end{tabular}
% \end{table*}
%
% \end{comment}

\begin{table*}[htbp]
\centering
\caption{Equal Error Rate (EER) obtained for each backbone using two approaches: the best single-layer with mean pooling (BSL), and multi-head factorized attentive pooling (MHFA). The best layer number for the BSL approach is indicated in the column "BSL \#", selected manually as an example to demonstrate the potential of identifying a relevant layer for each model. The column average corresponds to the average of the EER for the four corpora InTheWild, MLAAD, ASVspoof5-eval and LlamaPartialSpoof.}
\label{tab:MHFA_results_2}
%\vspace{-2mm}
\centering
\begin{tabular}{ll|rrrrrrrrrrrr}
\toprule
& & \multicolumn{2}{c}{ASVspoof5-dev} & \multicolumn{2}{c}{InTheWild} & \multicolumn{2}{c}{MLAAD} & \multicolumn{2}{c}{ASVspoof5-eval} & \multicolumn{2}{c}{LlamaPartialSpoof}  & \multicolumn{2}{c}{average} \\
\cmidrule(lr){3-4} \cmidrule(lr){5-6} \cmidrule(lr){7-8}\cmidrule(lr){9-10}\cmidrule(lr){11-12}\cmidrule(lr){13-14}
Backbone & BSL \# & BSL &  MHFA & BSL  & MHFA & BSL & MHFA& BSL  & MHFA & BSL  & MHFA & BSL  & MHFA \\
\midrule
BEATs          & 6 & 15.9 & 16.6 & 42.0 & 40.2 &  \textbf{4.1} &  7.7 & 31.8 & 35.5 & 18.5 & 18.3 & 24.1 & 25.4 \\
Wav2vec 2.0    & 8 &  6.4 &  5.9 & 17.9 & 16.8 & 14.3 & 20.2 & 20.3 & 12.1 & 30.5 & 23.7 & 20.7 & 18.2 \\
WavLM Base     & 4 &  6.9 &  2.5 & 17.8 & 16.7 & 29.0 & 16.1 &  9.5 &  6.9 & 31.7 & 21.1 & 22.0 & 15.2 \\
WavLM Large    & 8 &  \textbf{0.6} &  1.5 &  6.0 &  \textbf{5.4} & 10.5 & 15.9 &  \textbf{6.3} &  \textbf{5.1} & 21.3 & 27.5 &  11.0 & 13.5\\
MMS            & 9 &  2.2 &  6.2 & 17.0 & 18.1 &  8.5 & 11.4 & 10.5 &  7.4 & \textbf{13.3} & 28.3 & 12.3 & 13.8 \\
Wav2vec2-XLS-R & 6 &  1.9 &  \textbf{0.2} &  \textbf{5.6} &  9.8 & 11.9 &  \textbf{6.6} &  9.1 &  5.3 & 15.8 & \textbf{15.9} &  \textbf{10.6} &  \textbf{9.4} \\
\bottomrule
\end{tabular}
\end{table*}

\begin{table*}[htbp]
\centering
\caption{Equal Error Rate (EER) obtained with different fusion algorithms (for each line, there are two systems correspondig to fusion of BSL and fusion of MHFA systems). The fusion corpus is ASVspoof5-dev.}
\label{tab:MHFA_fusion}
%\vspace{-2mm}
\centering
\begin{tabular}{ll|rrrrrrrrrr}
\toprule
& & \multicolumn{2}{c}{InTheWild} & \multicolumn{2}{c}{MLAAD} & \multicolumn{2}{c}{ASVspoof5-eval} & \multicolumn{2}{c}{LlamaPartialSpoof}& \multicolumn{2}{c}{average}\\
\cmidrule(lr){3-4} \cmidrule(lr){5-6} \cmidrule(lr){7-9}  \cmidrule(lr){9-10} \cmidrule(lr){11-12} \\
Backbones & method & BSL & MHFA & BSL & MHFA & BSL &  MHFA & BSL &  MHFA & BSL &  MHFA\\
\midrule
BEATS, Wav2vec2.0, WavLM Base,  & fusion & 6.5 &  6.8 &  12.2 &  9.2 & 5.8 &   4.8 & 16.2 & 15.7 &  10.2 & 9.1 \\
WavLM Large, MMS and Wav2vec2-XLS-R  & sum &  7.1 & 7.2 &  6.8 & \textbf{5.4} &   5.6& \textbf{3.7} & 14.7 & 15.1 & 8.6& 7.9 \\
% all 6  & normalization and sum &   6.9 & 8.0 &  6.1 & 4.6 & 6.0 & 4.2 \\
\midrule
BEATS, WavLM Large,  & fusion & \textbf{5.4} & \textbf{6.3} & 9.7 & 6.4 & \textbf{5.5} & 4.7& 16.5 & 16.0 & 9.3& 8.3 \\
  MMS and Wav2vec2-XLS-R & sum &  7.7 & 6.4& \textbf{6.6} & 5.8& 6.4 & 3.9& \textbf{11.6}  & \textbf{14.7} & \textbf{8.1} & \textbf{7.7} \\
% wav2vec2-XLSR, BEATs, WavLM-large, MMS  & normalization and sum  &7.7  &  8.0& 6.3 & 4.7& 6.8 & 4.5&  \\
% \midrule
% wav2vec2-XLSR, BEATs & logistic regression & 5.6 & 9.7 & 12.1 & 6.3 & 9.0 & 5.4  \\
% wav2vec2-XLSR, BEATs & calibration and sum & 9.1 & 10.8& 6.7 &3.6 & 10.8 & 6.6 & \\
% wav2vec2-XLSR, BEATs & normalization and sum  & 10.6 & 13.8& 5.9 &2.9& 11.6 & 8.4\\
\bottomrule
\end{tabular}
\end{table*}

\subsection{Performance of individual models on OOD conditions}

The results of the layer-wise analysis are shown in \autoref{fig:layer_wise_analysis}, where we illustrate the evolution of the EER as a function of the layer. The general trend is consistent across different backbones, with performance improving through intermediate layers before degrading in deeper layers.

For each model, we manually select the best single-layer (BSL) for each model as an intermediate layer with best performance across OOD corpora. The BSL can be considered as an oracle system to evaluate the potential of single layer models with optimal choice of the layer.

\autoref{tab:MHFA_results_2} presents the EER obtained for each backbone with the best single-layer (BSL) model and MHFA, on five OOD corpora. The last column corresponds to the average of the EER on the four corpora InTheWild, MLAAD+M-AILABS, ASVspoof5-eval and LlamaPartialSpoof.
Despite using all transformer layers, MHFA does not consistently outperform the oracle BSL model across all models and datasets, with results varying depending on the backbone and data characteristics.

% The generalization of detectors to OOD conditions depends highly on the SSL model with similar trends for BSL and MHFA systems. It is not surprising that wav2vec2-XLS-R, MMS and WavLM-large perform better on the multilingual MLAAD corpus than their monolingual counterparts. More surprising is the very good performance of BEATs on MLAAD despite poor generalization on ASVspoof5, this behavior would necessitate more analysis to understand whehter its non-speech based pretraining objective is complementary to speech-based SSL models for audio deepfake detection. The detection of partially generated utterances on LlamaPartialSpoof is very challenging for all models trained on only fully fake utterances.

%
% While MHFA outperforms LWA for BEATs, Wav2Vec 2.0, and WavLM Base, it performs worse on some datasets for WavLM Large and MMS, suggesting that its impact is not universally beneficial.
%
% We analyzed the evolution of the weights for all backbones and both LWA and MHFA heads. In each case, the observed trends were consistent across different architectures and weight types. Therefore, \autoref{fig:layer_weight_analysis} only presents the evolution of the weights $w^{v}$ for WavLM Base and WavLM at different training checkpoints.
% The weight distribution remains flat throughout training, indicating that the results are essentially an average across all layers. The slight prioritization of layers 1 to 5 is negligible in absolute value.

\subsection{Contribution of each layer}

The layer weights in LWA classfication heads can be interpreted as contributions of each layer to the deepfake detection task, as proposed by
\cite{kheir_comprehensive_2025}.  \cite{kheir_comprehensive_2025} concludes that all models follow a similar trend with the first layers (4-6 layers for small models and 10-12 layers for larger models) being the most important for deepfake detection. Following this approach, we plot in Figure \ref{fig:MHFA_weights} the weights $w^k$ and $w^v$ of the MHFA heads trained for each SSL model. Since the weights sum to one for each model, we use a different color scale for each size of model.

We observe a moderate variation of the weights between the different layers. The trends seem similar for the key and value weights. For the models Wav2vec 2.0, WavLM Base, WavLM Large and MMS, the prioritization of the layers observed in MHFA weights seem consistent with the best single layer selection. Conversely, BEATs and Wav2vec2 XLSR  have important MHFA weights for high layers of the model.

% We provide more experimental support to their conclusion with two contributions.
% \begin{enumerate}
%  \item The performance of classifiers trained on each layer gives the same trend as the analysis of LWA weights.
%  \item We show that the results hold for out-of-domain generalization, extending their analysis for in-domain performance.
% \end{enumerate}
%

\subsection{Performance of fusion algorithms}

\autoref{tab:MHFA_fusion} presents the performance of the fusion of systems trained for the six SSL encoders, and the fusion of four complementary models. For each set of backbone models and fusion method, two systems are evaluated, corresponding to the fusion of BSL models and to the fusion of MHFA models.

The fusion method has an impact on the performance, the logistic regression performing better on ITW and worst on MLAAD than the sum of calibrated scores, maybe because of the closer proximity with the fusion corpus. Fusion methods achieve a balanced performance across OOD corpora, for instance with the fusion of the MHFA systems wav2vec2-XLS-R, BEATS, MMS and WavLM Large achieving an EER of 6.4\% on ITW, 5.8 \% on MLAAD and 3.9 \% on ASVspoof5-eval.

% \subsection{Comparison to state-of-the-art performance}

\subsection{Discussion}

\subsubsection{Layer-wise analysis}

All models exhibit a similar pattern: the error rate is initially high, decreases sharply through intermediate layers, and then increases towards the output layer (\autoref{fig:layer_wise_analysis}). This pattern is consistent across all datasets, emphasizing the importance of intermediate layers for deepfake detection. Relying solely on the output layer may not be the most effective approach.
Overall, the layers following the initial sharp decline in EER strike a good balance in performance across different datasets.

\subsubsection{Effectiveness of the layer weighted average pooling}

When analyzing the evolution of layer weights in terms of relative behavior, we observe that earlier layers are slightly favored for four backbones over six, which aligns with the findings in \cite{kheir_comprehensive_2025}. In our experiments, we observe that these early layers correspond to the best generalization performance of single layer models.%  However this observation is not consistent for all beckbones, motivating the layer wise analysis to select %However, the differences between layers appear small in terms of absolute contributions for both LWA and MHFA regarding their impact on the classification task. As a result, the model's behavior tends to approach an average, with layer contributions appearing relatively uniform.
% This could be due to the simplicity of the training task, as we observe a rapid convergence to near-zero validation loss.
% Because the layer impacts are not distinguishable, unlike what we observe in the layer-wise analysis, we suggest that LWA weights may not be the most suitable for discussing the information extracted from SSL for deepfake detection.

MHFA achieves performance comparable or superior to the best single-layer approach without the need to explicitly identify the most relevant layer. Nevertheless the selection of a single layer allow an early exit strategy, as suggested by \cite{pimentel2024efficient}. The fusion experiments suggest that with a given computational budget, it may be more efficient to use intermediate layers of several SSL models than using all layers of a single encoder.

\subsubsection{Out-of-domain performance variability across SSL encoders}

Performance varies significantly across models for a given dataset, with similar trends for BSL and MHFA. For instance, WavLM Large achieves an EER of 5.4\% on ITW, while MMS struggles with an EER exceeding 10\%. Increasing model size does not necessarily lead to better performance.
This suggests that some backbones may exhibit variable generalization capabilities due to their pre-training objectives and datasets.
For example, WavLM was explicitly trained with a strong emphasis on noise robustness \cite{chen_wavlm_2022}, which could contribute to its superior performance.

 It is not surprising that Wav2vec2.0 XLS-R, MMS and WavLM Large perform better on the multilingual MLAAD corpus than their monolingual counterparts.  More surprising is BEATs, which achieves an EER of 4.1\%.
However, BEATs performs much worse on ITW compared to WavLM Large. This suggests that OOD performance for SSL encoders depends on the encoder itself, potentially due to differences in training objectives.
BEATs, trained as a general audio model rather than being specialized solely on speech, might rely more on speaker-independent features \cite{hai_ghost--wave_2024}.

\subsubsection{Fusion of simple models}

The fusion of four simple classifiers trained with frozen backbones with limited training data and no data augmentation achieve a competitive generalization performance on several OOD corpora This illustrates the complementarity of different SSL models to detect audio deepfakes on OOD conditions, as suggested by \cite{kulkarni2024exploring}. %, as compared to \cite{kulkarni2024exploring, falez2024whispeak}.

\subsection{Limitations}

Our analysis is limited to a simple training recipe, with the ASVspoof5 train and dev as training and calibration corpus, no data augmentation and no finetuning of the SSL backbones. Future work should evaluate the validity of our conclusions for more sophisticated training recipes.
\begin{itemize}
 \item The use of larger training datasets, including other languages, with data augmentation strategies, is expected to improve performance on unseen conditions.
 \item The finetuning of the SSL encoder with the deepfake detection objective could also affect our conclusions. The finetuning with specific attack could improve performance on in-domain conditions with the risk of degrading generalization to OOD conditions.
\end{itemize}

\section{Conclusion}
\label{sec:section5}

%JF
In this work, we studied the capabilities of the main pre-trained SSL encoders in the audio deepfakes detection task, when these encoders are frozen (not tuned for the task).
The level of performance achieved using the appropriate frozen encoder for a specific corpus is state-of-the-art, with an EER of 5.4\% on InTheWild with WavLM Large, for example. 
We also observed a strong variation in performance depending on the encoder selected and the corpus targeted. 
In particular, BEATs, a general audio model, excelled on the MLAAD dataset with a minimum EER of 4.1\% (to be compared with EER ranging from 8.5\% to 29\% for the others), while it was on of the lowest in terms of performance on the other corpora. 

We also looked at the information embedded at the different layers of the encoders. 
For all models, intermediate layers have proved indispensable.
Our experiments have also shown that connecting a classification head directly to the output layer is detrimental to performance.

We have compared different classification head architectures.
It is interesting to note that a simple classification head, such as average pooling, plugged into the most relevant layer, can compete with a more complex method such as MHFA, with a significantly lower computational cost.

Finally, we sought to combine the different encoders to improve OOD deepfake detection capabilities. 
We found that a simple score-level fusion of four frozen-SSL-based systems achieved the best generalization performance for a large range of OOD conditions.

This work confirms the value of pre-trained frozen SSLs for building powerful deepfake detection systems at a fairly low level of effort.
Two limitations of using frozen SSL models for deepfake detection should be noted here. Firstly, the variation in performance as a function of SSL model and test set requires a fuller and deeper understanding of model behavior before moving on to the real world.
Secondly, it seems obvious that an attacker can use knowledge of the nature of the system, the use of a frozen, pre-trained SSL encoder, in the design of the attack.
In future works, we will adress these two limitations.

\section*{Acknowledgment}

This work was performed using HPC resources from GENCI–IDRIS (Grant 2025-AD011014982R1).

\bibliographystyle{IEEEtran}
\bibliography{Deepfake_2025}

\begin{thebibliography}{10}
\providecommand{\url}[1]{#1}
\csname url@samestyle\endcsname
\providecommand{\newblock}{\relax}
\providecommand{\bibinfo}[2]{#2}
\providecommand{\BIBentrySTDinterwordspacing}{\spaceskip=0pt\relax}
\providecommand{\BIBentryALTinterwordstretchfactor}{4}
\providecommand{\BIBentryALTinterwordspacing}{\spaceskip=\fontdimen2\font plus
\BIBentryALTinterwordstretchfactor\fontdimen3\font minus
  \fontdimen4\font\relax}
\providecommand{\BIBforeignlanguage}[2]{{%
\expandafter\ifx\csname l@#1\endcsname\relax
\typeout{** WARNING: IEEEtran.bst: No hyphenation pattern has been}%
\typeout{** loaded for the language `#1'. Using the pattern for}%
\typeout{** the default language instead.}%
\else
\language=\csname l@#1\endcsname
\fi
#2}}
\providecommand{\BIBdecl}{\relax}
\BIBdecl

\bibitem{wang24_ASVspoof}
X.~Wang, H.~Delgado, H.~Tak, J.~weon Jung, H.~jin Shim, M.~Todisco, I.~Kukanov,
  X.~Liu, M.~Sahidullah, T.~H. Kinnunen, N.~Evans, K.~A. Lee, and J.~Yamagishi,
  ``Asvspoof 5: crowdsourced speech data, deepfakes, and adversarial attacks at
  scale,'' in \emph{The Automatic Speaker Verification Spoofing Countermeasures
  Workshop (ASVspoof 2024)}, 2024, pp. 1--8.

\bibitem{muller_does_2024}
N.~Müller, P.~Czempin, F.~Diekmann, A.~Froghyar, and K.~Böttinger, ``Does
  audio deepfake detection generalize?'' in \emph{Interspeech 2022}.\hskip 1em
  plus 0.5em minus 0.4em\relax {ISCA}, 2022, pp. 2783--2787.

\bibitem{doan_trident_2024}
\BIBentryALTinterwordspacing
T.-P. Doan, H.~Dinh-Xuan, T.~Ryu, I.~Kim, W.~Lee, K.~Hong, and S.~Jung,
  ``Trident of poseidon: A generalized approach for detecting deepfake
  voices,'' in \emph{Proceedings of the 2024 on {ACM} {SIGSAC} Conference on
  Computer and Communications Security}.\hskip 1em plus 0.5em minus 0.4em\relax
  {ACM}, pp. 2222--2235.
\BIBentrySTDinterwordspacing

\bibitem{zhang_audio_2024}
\BIBentryALTinterwordspacing
Q.~Zhang, S.~Wen, and T.~Hu, ``Audio deepfake detection with self-supervised
  {XLS}-r and {SLS} classifier,'' in \emph{Proceedings of the 32nd {ACM}
  International Conference on Multimedia}.\hskip 1em plus 0.5em minus
  0.4em\relax {ACM}, pp. 6765--6773.
\BIBentrySTDinterwordspacing

\bibitem{tak2022automatic}
H.~Tak, M.~Todisco, X.~Wang, J.-w. Jung, J.~Yamagishi, and N.~W. Evans,
  ``Automatic speaker verification spoofing and deepfake detection using
  wav2vec 2.0 and data augmentation,'' in \emph{Odyssey}, 2022.

\bibitem{martin-donas_exploring_2024}
\BIBentryALTinterwordspacing
J.~M. Martín-Doñas, A.~Álvarez, E.~Rosello, A.~M. Gomez, and A.~M. Peinado,
  ``Exploring self-supervised embeddings and synthetic data augmentation for
  robust audio deepfake detection,'' in \emph{Interspeech 2024}.\hskip 1em plus
  0.5em minus 0.4em\relax {ISCA}, pp. 2085--2089.
\BIBentrySTDinterwordspacing

\bibitem{pascu2024towards}
O.~Pascu, A.~Stan, D.~Oneata, E.~Oneata, and H.~Cucu, ``Towards generalisable
  and calibrated audio deepfake detection with self-supervised
  representations,'' in \emph{Interspeech}, vol. 2024, 2024, pp. 4828--4832.

\bibitem{kulkarni2024exploring}
A.~Kulkarni, H.~M. Tran, A.~Kulkarni, S.~Dowerah, D.~Lolive, and M.~M. Doss,
  ``Exploring generalization to unseen audio data for spoofing: Insights from
  ssl models,'' in \emph{ASVSpoof workshop 2024}, 2024.

\bibitem{kheir_comprehensive_2025}
\BIBentryALTinterwordspacing
Y.~E. Kheir, Y.~Samih, S.~Maharjan, T.~Polzehl, and S.~Möller, ``Comprehensive
  layer-wise analysis of ssl models for audio deepfake detection,'' 2025,
  accepted to NAACL Findings 2025.
\BIBentrySTDinterwordspacing

\bibitem{pimentel2024efficient}
A.~Pimentel, Y.~Zhu, H.~R. Guimar{\~a}es, and T.~H. Falk, ``Efficient audio
  deepfake detection using wavlm with early exiting,'' in \emph{2024 IEEE
  International Workshop on Information Forensics and Security (WIFS)}.\hskip
  1em plus 0.5em minus 0.4em\relax IEEE, 2024, pp. 1--6.

\bibitem{pasad_layer-wise_2022}
\BIBentryALTinterwordspacing
A.~Pasad, J.-C. Chou, and K.~Livescu, ``Layer-wise analysis of a
  self-supervised speech representation model,'' \emph{IEEE Automatic Speech
  Recognition and Understanding Workshop - ASRU 2021}.
\BIBentrySTDinterwordspacing

\bibitem{peng_attention-based_2023}
\BIBentryALTinterwordspacing
J.~Peng, O.~Plchot, T.~Stafylakis, L.~Mosner, L.~Burget, and J.~Cernocky, ``An
  attention-based backend allowing efficient fine-tuning of transformer models
  for speaker verification,'' in \emph{2022 {IEEE} Spoken Language Technology
  Workshop ({SLT})}.\hskip 1em plus 0.5em minus 0.4em\relax {IEEE}, pp.
  555--562.
\BIBentrySTDinterwordspacing

\bibitem{baevski_wav2vec_2020}
\BIBentryALTinterwordspacing
A.~Baevski, Y.~Zhou, A.~Mohamed, and M.~Auli, ``wav2vec 2.0: A framework for
  self-supervised learning of speech representations,'' in \emph{Advances in
  Neural Information Processing Systems}, vol.~33.\hskip 1em plus 0.5em minus
  0.4em\relax Curran Associates, Inc., pp. 12\,449--12\,460.
\BIBentrySTDinterwordspacing

\bibitem{chen_wavlm_2022}
\BIBentryALTinterwordspacing
S.~Chen, C.~Wang, Z.~Chen, Y.~Wu, S.~Liu, Z.~Chen, J.~Li, N.~Kanda,
  T.~Yoshioka, X.~Xiao, J.~Wu, L.~Zhou, S.~Ren, Y.~Qian, Y.~Qian, J.~Wu,
  M.~Zeng, X.~Yu, and F.~Wei, ``{WavLM}: Large-scale self-supervised
  pre-training for full stack speech processing,'' vol.~16, no.~6, pp.
  1505--1518, {IEEE} Journal of Selected Topics in Signal Processing.
\BIBentrySTDinterwordspacing

\bibitem{pratap_scaling_2023}
\BIBentryALTinterwordspacing
V.~Pratap, A.~Tjandra, B.~Shi, P.~Tomasello, A.~Babu, S.~Kundu, A.~Elkahky,
  Z.~Ni, A.~Vyas, M.~Fazel-Zarandi, A.~Baevski, Y.~Adi, X.~Zhang, W.-N. Hsu,
  A.~Conneau, and M.~Auli, ``Scaling speech technology to 1,000+ languages,''
  \emph{Journal of Machine Learning Research}, vol.~25, no.~97, pp. 1--52,
  2024.
\BIBentrySTDinterwordspacing

\bibitem{babu2022xls}
A.~Babu, C.~Wang, A.~Tjandra, K.~Lakhotia, Q.~Xu, N.~Goyal, K.~Singh, P.~von
  Platen, Y.~Saraf, J.~Pino \emph{et~al.}, ``Xls-r: Self-supervised
  cross-lingual speech representation learning at scale,'' in \emph{Proc.
  Interspeech 2022}, 2022, pp. 2278--2282.

\bibitem{zaiem_speech_2023}
\BIBentryALTinterwordspacing
S.~Zaiem, Y.~Kemiche, T.~Parcollet, S.~Essid, and M.~Ravanelli, ``Speech
  self-supervised representations benchmarking: A case for larger probing
  heads,'' \emph{Computer Speech \& Language}, vol.~89, p. 101695, 2025.
\BIBentrySTDinterwordspacing

\bibitem{chen_beats_2022}
\BIBentryALTinterwordspacing
S.~Chen, Y.~Wu, C.~Wang, S.~Liu, D.~Tompkins, Z.~Chen, W.~Che, X.~Yu, and
  F.~Wei, ``{BEAT}s: Audio pre-training with acoustic tokenizers,'' in
  \emph{Proceedings of the 40th International Conference on Machine Learning},
  ser. Proceedings of Machine Learning Research, A.~Krause, E.~Brunskill,
  K.~Cho, B.~Engelhardt, S.~Sabato, and J.~Scarlett, Eds., vol. 202.\hskip 1em
  plus 0.5em minus 0.4em\relax PMLR, 23--29 Jul 2023, pp. 5178--5193.
\BIBentrySTDinterwordspacing

\bibitem{pianese_training-free_2024}
\BIBentryALTinterwordspacing
A.~Pianese, D.~Cozzolino, G.~Poggi, and L.~Verdoliva, ``Training-free deepfake
  voice recognition by leveraging large-scale pre-trained models,'' in
  \emph{Proceedings of the 2024 {ACM} Workshop on Information Hiding and
  Multimedia Security}, pp. 289--294.
\BIBentrySTDinterwordspacing

\bibitem{muller_mlaad_2024}
N.~M. Müller, P.~Kawa, W.~H. Choong, E.~Casanova, E.~Gölge, T.~Müller,
  P.~Syga, P.~Sperl, and K.~Böttinger, ``Mlaad: The multi-language audio
  anti-spoofing dataset,'' in \emph{2024 International Joint Conference on
  Neural Networks (IJCNN)}, 2024, pp. 1--7.

\bibitem{luong2025llamapartialspoof}
H.-T. Luong, H.~Li, L.~Zhang, K.~A. Lee, and E.~S. Chng, ``Llamapartialspoof:
  An llm-driven fake speech dataset simulating disinformation generation,'' in
  \emph{ICASSP 2025-2025 IEEE International Conference on Acoustics, Speech and
  Signal Processing (ICASSP)}.\hskip 1em plus 0.5em minus 0.4em\relax IEEE,
  2025, pp. 1--5.

\bibitem{alumae2019taltech}
A.~Tanel~Alum\"{a}e, ``The {TalTech} systems for the {VOiCES from a Distance
  Challenge},'' in \emph{Interspeech}, 2019, pp. 746--750.

\bibitem{hai_ghost--wave_2024}
\BIBentryALTinterwordspacing
X.~Hai, X.~Liu, Z.~Chen, Y.~Tan, S.~Li, W.~Niu, G.~Liu, R.~Zhou, and Q.~Zhou,
  ``Ghost-in-wave: How speaker-irrelative features interfere deepfake voice
  detectors,'' in \emph{2024 {IEEE} International Conference on Multimedia and
  Expo ({ICME})}.\hskip 1em plus 0.5em minus 0.4em\relax {IEEE}, pp. 1--6.
\BIBentrySTDinterwordspacing

\end{thebibliography}

% \vspace{12pt}
% \color{red}
% IEEE conference templates contain guidance text for composing and formatting conference papers. Please ensure that all template text is removed from your conference paper prior to submission to the conference. Failure to remove the template text from your paper may result in your paper not being published.

\end{document}